\documentclass{aastex}
\usepackage{emulateapj5}
\usepackage{apjfonts}
\usepackage{epsfig}

\newcommand{\PSbox}[3]{\mbox{\rule{0in}{#3}\includegraphics{#1}\hspace{#2}}}

\newcommand{\kms}{~{\rm km~s^{-1}}}
\newcommand{\msun}{~M_\odot}

\newcommand{\yr}{~{\rm yr}}
\newcommand{\cm}{~{\rm cm}}
\newcommand{\pc}{~{\rm pc}}
\newcommand{\kpc}{~{\rm kpc}}	
\newcommand{\ergs}{~{\rm erg~s^{-1}}}
\newcommand{\kev}{~{\rm keV}}
\newcommand{\ev}{~{\rm eV}}
\newcommand{\lx}{L_{\rm X}}
\newcommand{\cs}{c_{\rm s}}

\newcommand{\mdot}{~M_\odot~{\rm yr}^{-1}}

\shorttitle{DIM X-RAY SOURCES IN GLOBULAR CLUSTERS}
\shortauthors{PFAHL \& RAPPAPORT}

\begin{document}


\submitted{Accepted for Publication in ApJ}

\title{Bondi-Hoyle-Lyttleton Accretion Model for Low-luminosity X-ray Sources in Globular Clusters}

\author{Eric Pfahl and Saul Rappaport}
\affil{Department of Physics, Massachusetts Institute of Technology, Cambridge,
MA, 02139 \\ pfahl@space.mit.edu, sar@mit.edu}


\begin{abstract}We present a new model for low-luminosity X-ray sources in globular clusters, with 
$\lx \lesssim 10^{34} \ergs$.  The model we propose is that of a single neutron star accreting 
from cluster gas that has accumulated as a natural product of stellar evolution.  An analytic luminosity 
function is derived under the assumption that the speed distribution of neutron stars in the 
central region of a cluster is described by a Maxwellian, and that the density and temperature of the 
gas are uniform.  Predictions of the model and implications for the gas content of 
globular clusters are discussed.
\end{abstract}




\section{INTRODUCTION}

Globular clusters contain at least two varieties of X-ray sources, 
differentiated by their luminosities.  There are a dozen bright X-ray sources observed 
in the Milky Way globular cluster system, with $\lx\sim10^{36}-10^{38}\ergs$ \citep{deutsch00}.  
Nearly all of these higher luminosity sources show type I X-ray bursts \citep{lewin93}, and it is, 
therefore, clear that they are accreting neutron stars (NSs) in binary systems.  In fact, 7 of the 
12 bright sources have well-measured or constrained binary periods \citep{deutsch00}.  
There is also a population of dim cluster X-ray sources (DCXSs), with $\lx\sim10^{31}-10^{34}\ergs$, 
where the lower limit is set by the detection sensitivities. 
To date, $\gtrsim30$ of these dim sources have been observed,
primarily with the {\em ROSAT} and {\em Einstein} satellites \citep{hertz83a,hertz83b,rappaport94,johnston96a}. 
The nature of the DCXSs remains a mystery, largely 
due to the fact that sufficiently accurate positions have not been available to allow for unambiguous 
optical or radio identifications in the host clusters.

In a campaign to search for cataclysmic variables (CVs) in globular clusters, {\em HST} observations
have discovered {\em possible} optical counterparts to several of the DCXSs in the clusters surveyed 
\citep{grindlay93,cool95}.  The absolute position accuracy of the {\em HST} counterparts is  
$\lesssim 0\farcs5$, while the {\em ROSAT} positions of the DCXSs are no better than $\sim 2''$ 
\citep{verbunt98}.  In addition, the DCXSs are concentrated toward the centers of their respective clusters, 
where the stellar density is high.  High stellar density and source concentration, coupled with the disparity 
between {\em HST} and {\em ROSAT} position accuracies, make source confusion a serious issue.  
Current and pending {\em Chandra} observations are likely to resolve the problem of source identification.  

As a population, the DCXSs are characterized by their low luminosities, temperatures 
$kT \sim 0.1 - 0.5 \kev$ \citep{johnston96a} for an assumed blackbody spectrum, and the fact that they 
typically lie within a few core radii from the centers of their host clusters 
\citep{johnston96b,verbunt98}.  However, this information provides only weak constraints for models 
of the DCXSs.  The hypothesis that the DCXSs are CVs \citep{hertz83a} is appealing, but the X-ray luminosities
of CVs in the Galactic disk typically lie in the range $\lx \sim 10^{30} - 10^{32.5}$ 
\citep[see Fig. 8 in][]{verbunt97}, falling short of the higher luminosities observed 
among the DCXSs \citep{verbunt84}.   Other suggestions regarding the nature of the DCXSs 
include low-mass X-ray binaries in quiescence \citep{verbunt84}, and rapidly rotating NSs
that have not yet been detected as millisecond radio pulsars (see Becker \& Tr\"umper 1999 for X-ray 
luminosities of known millisecond radio pulsars).  In this work we propose an alternate model 
that may explain some fraction of the DCXSs. 

A NS moving through a gaseous medium
will be able to accrete some of this material via the Bondi-Hoyle-Lyttleton (BHL) process 
\citep{hoyle41,bondi44,bondi52}, 
converting a fraction of the gravitational energy of the
accreted gas into X-ray radiation.  This physical picture was first proposed by \citet{ostriker70}, 
and later extended by \citet{treves91} and \citet{blaes93}, for NSs in the Galactic
disk, where the NS is assumed to accrete from the interstellar medium or from the gas in a giant 
molecular cloud.  We propose the application of this basic idea to globular clusters to explain the 
DCXSs.  This model has the rather
minimal requirements that (i) there be a population of single NSs able to accrete from the intracluster gas, 
and (ii) the ambient gas density and temperature lie in the right range to yield the requisite 
accretion luminosities.

One major point of uncertainty regarding BHL accretion by the Galactic {\em disk} 
population of single NSs is their distribution
of space velocities \citep[see][]{hansen97}.  If the NS is moving supersonically through a 
gas, with speed $v$, then the BHL accretion luminosity is proportional to $v^{-3}$, implying that an 
uncertainty in the speed 
distribution is considerably amplified when computing the luminosity function of the accreting
sources.  The extension of the BHL accretion model to NSs in globular clusters does not suffer 
from this indeterminacy.  
For the more centrally concentrated globular clusters (e.g., 47 Tuc, M15, NGC 6397) the central relaxation
time is typically $\lesssim 8~{\rm Gyr}$, so that at the current epoch the stars in the cluster core
region should be nearly in thermal equilibrium \citep[see][]{watters00} 
and the speed distribution for objects 
of a certain mass should be roughly Maxwellian for speeds less than the cluster escape speed.
With a well-defined speed distribution, it is straightforward to compute the shape of the 
luminosity function for the isolated, accreting cluster NSs (IACNs).  
In this regard, a model qualitatively similar to our own (using a Maxwellian speed distribution), 
but applied to the Galactic center, was proposed by \citet{zane96} to explain the diffuse X-ray 
emission from the Galactic center.  

In \S \ref{sec:acc} we compute the accretion luminosity for a NS moving through a gaseous medium.  
We discuss the range of intracluster gas density and temperature required to explain the DCXSs.  
In \S \ref{sec:lum} the luminosity function for the IACNs is derived under the assumptions of 
a Maxwellian distribution in NS speeds, and a constant gas density and temperature over 
the region of interest.  Caveats concerning our model are discussed in \S \ref{sec:dis}.     
Finally, in \S \ref{sec:con} we summarize some of the fundamental points of the IACN model. 


\section{ACCRETION FROM THE INTRACLUSTER GAS}\label{sec:acc}

A NS of mass $M_{\rm ns}$, moving with relative speed $v$ through a gas of ambient density $\rho$ and 
sound speed $\cs$, accretes at the BHL rate \citep[see][]{foglizzo97},
\begin{equation}\label{eq:mdot}
{\dot M} \simeq 4 \pi (G M_{\rm ns})^2 \rho (v^2 + \cs^2)^{-3/2}~.
\end{equation}
If we define $V \equiv (v^2 + \cs^2)^{1/2}$ and $\rho \equiv m_{\rm p} n$, where $m_{\rm p}$ is the proton 
mass and $n$ is the hydrogen number density, then for a NS of mass $M_{\rm ns}=1.4\msun$ 
and radius $R_{\rm ns}=10~{\rm km}$, the corresponding X-ray luminosity is given by
\begin{eqnarray}\label{eq:acc-lum}
\lx & = & \epsilon G M_{\rm ns} {\dot M} R_{\rm ns}^{-1} \nonumber \\
& \simeq &  
10^{32} ~ \epsilon n \left( \frac{V}{10 \kms} \right)^{-3} \ergs~,
\end{eqnarray}
where $\epsilon \lesssim 1$ is the efficiency for converting gravitational energy into X-ray radiation,
and the density is in units of $\cm^{-3}$.

In order for the IACN model to successfully describe the DCXSs, the predicted X-ray luminosity must 
naturally span the range of the observed luminosities.  This requires that the resultant speed $V$
not be too large, and that the gas density not be too small.  
For those globular clusters where it is possible to measure the radial velocity dispersion, $\sigma$, of
the central population of stars, one finds $\sigma \sim 5-20 \kms$ \citep{dubath97}.
Generally, it is the light from giants, with mass $\sim 0.8 \msun$, that dominates the
sample used in the velocity dispersion measurements.  
Energy equipartition implies that the one-dimensional velocity dispersions
$\sigma_{\rm m1}$ and $\sigma_{\rm m2}$, for objects of mass $m_1$ and $m_2$, are related by 
$m_1 \sigma_{\rm m1}^2 = m_2 \sigma_{\rm m2}^2$ \citep{binney87}.  If the dynamical temperature 
of the NS population is the same as that of the turn-off stars with measured velocity dispersions, 
it follows that the one-dimensional velocity dispersion of NSs is $\sigma_{\rm ns} \lesssim 15 \kms$,
so that the speed itself should not present a problem for the IACN model of the DCXSs. 
The density of the gas and its thermal speed are more uncertain, and these uncertainties are tied to 
the larger problem of the gas content in globular clusters, which we now discuss. 

A star with the cluster turn-off mass ($\sim 0.8 \msun$) sheds most of its $\sim 0.2 \msun$ envelope as 
it ascends the AGB, with a wind outflow speed of $\sim 10-20 \kms$ \citep{knapp85}.  
Therefore, if the central escape speed of a globular cluster is $\gtrsim 20 \kms$, mass should 
accumulate in the central regions of the cluster.  For a cluster that contains a mass $M_{\rm c}$ of 
stars within a few core radii from its center, the expected {\rm total} rate of mass-loss from stars is
${\dot M}_{\rm wind} \sim 10^{-6}(M_{\rm c}/10^5 \msun)\mdot$ \citep[see][]{knapp73,tayler75}.
As a globular cluster passes through the midplane of the Galaxy, it is expected that any accumulated gas
will be swept out by ram pressure \citep{tayler75}.  In the $\sim 10^8$ years between midplane crossings, 
a globular cluster with $M_{\rm c} = 10^5 \msun$ could thereby accumulate of order $100 \msun$ of gas.  
This estimate is reduced considerably (by more than a factor of 10 in some cases) in more sophisticated 
treatments of the gas-flow problem in globular clusters which include heating and cooling processes 
\citep{scott75,vandenberg77,knapp96}.  

Direct observational searches for 
cluster gas in the form of molecular, neutral, and ionized hydrogen have yielded only non-detections 
\citep{smith90,smith95,knapp96}, implying upper limits to the total gas content in the range 
$\sim 0.1-10 \msun$.   In a search for ionized hydrogen in 
six globular clusters, \citet{knapp96} found $M_{\rm H_+} \lesssim 0.1 \msun$ within about one core
radius for the clusters observed, implying upper limits $n_{\rm H_+} \sim 50 - 100 \cm^{-3}$.   
A density of $100 \cm^{-3}$, however, gives a BHL accretion luminosity of $10^{34} \ergs$, which is 
certainly high enough to explain the brightest of the DCXSs.  

A simplistic argument can be made to determine a lower limit to the density of the intracluster gas.
Suppose that the wind from each star in the central region of a globular cluster is
allowed to flow freely, unimpeded by the gravity of the cluster and interactions with the wind
from other stars.  Furthermore, assume that the spatial distribution of stars is uniform and 
confined within a sphere of radius $r_\star$.  In this approximation, it can be shown that
for $r < r_\star$
\begin{eqnarray}\label{eq:den}
n & > & \frac{ {3\dot M}_{\rm wind} }{ 8 \pi r_\star^2 v_{\rm wind} m_{\rm p}} \nonumber \\
& \sim & 1 ~ \left( \frac{M_{\rm c}}{10^5 \msun} \right) 
\left( \frac{v_{\rm wind}}{20 \kms} \right)^{-1} 
\left( \frac{r_\star}{0.5 \pc} \right)^{-2}\cm^{-3} ~,
\end{eqnarray}
where the scale $r_\star=0.5 \pc$ has been chosen to represent the characteristic radius of a dense
globular cluster core.
When the cluster gravity and the interaction between stellar winds is taken into account, we suspect that the 
gas density can be substantially larger than this value, although a hydrodynamic 
investigation is certainly merited.  

At least two lines of argument suggest that the temperature of the intracluster gas should be 
$\gtrsim 10^4~{\rm K}$, with a sound speed $\cs \sim 10 ~ (T/10^4~{\rm K})^{1/2} \kms$.  Only one or two
hot, post-AGB stars are required to provide sufficient ultraviolet flux to leave the  
gas near the cluster center in a warm ($T \sim 10^4~{\rm K }$), photoionized state 
\citep[see][]{osterbrock89,knapp96}.  However, a simple calculation shows that only of order one in ten 
clusters should contain a hot, post-AGB star, owing to the short lifetime ($\lesssim 10^5 \yr$) of the
phase.  We return to this point in \S \ref{sec:dis} and discuss the consequences of a 
predominantly neutral intracluster gas for the IACN model. In addition to discrete ionizing energy 
sources within the cluster, complete thermalization of the outflow speeds of $\sim 10-20 \kms$, via 
collisions between stellar winds, would also translate to a temperature of $\sim 10^4~{\rm K}$.

We proceed under the assumption that there are no processes whereby the gas is continuously heated 
so that it maintains a very high temperature ($\gtrsim 10^5~{\rm K}$), or is removed so efficiently as to 
render moot the discussion of the intracluster gas and the IACN model.  A number of energetic gas 
removal mechanisms have been suggested \citep[for a list, see][]{smith95}.  Of these mechanisms, the 
most powerful is perhaps the sweeping/heating of the intracluster gas by relativistic winds or 
low-frequency radiation from millisecond pulsars \citep{spergel91}. Millisecond pulsars are known to 
be abundant in some globular clusters \citep[see][]{camilo00}, but there is no direct evidence
yet that the pulsar wind sweeping mechanism is actually operative.


\section{THE LUMINOSITY FUNCTION}\label{sec:lum}

The derivation of the luminosity function (LF) is simplified considerably if the assumption is made that
the gas density, $n$, the sound speed, $\cs$, and the one-dimensional velocity dispersion of NSs, 
$\sigma_{\rm ns}$, are constant over the region of interest.  
These assumptions should be reasonable within the core region of a globular cluster.
Given the NS speed distribution, $p(v)$, the X-ray LF is obtained
by computing $p(L)=|dv/dL|p(v)$, where we have dropped the subscript on $\lx$.  We assume that 
the speed distribution in the central region of a globular cluster is well-represented by a Maxwellian:
\begin{equation}\label{eq:max}
p(v)dv = \sqrt{ \frac{2}{\pi} } \frac{v^2}{\sigma_{\rm ns}^3} 
\exp \left(-\case{1}{2}v^2/\sigma_{\rm ns}^2 \right) dv~.
\end{equation}
This is certainly an idealization, since a more realistic speed distribution would vanish beyond the
cluster escape speed.  However, we would like to stress that the LF derived here is for illustrative 
purposes only.  There are a number of important processes which may lead to a LF that is quite different
from the one given below (see \S \ref{sec:dis}).   

For convenience, we define the following dimensionless quantities:
\begin{equation}\label{eq:dimless}
\ell \equiv L/L_{\rm max}~; 
~{\cal M} \equiv v/\cs ~; 
~\mu \equiv \sigma_{\rm ns}/\cs~,
\end{equation}
where $L_{\rm max} \equiv 4 \pi \epsilon (G M_{\rm ns})^3 R_{\rm ns}^{-1} \rho \cs^{-3}$ is the 
maximum accretion luminosity (eq. [\ref{eq:acc-lum}], with $v=0$).
In terms of the Mach number ${\cal M}(\ell) = (\ell^{-2/3} - 1)^{1/2}$, the LF
is 
\begin{equation}\label{eq:lf}
p(\ell)d\ell = 
\mu^{-3} \sqrt{ \frac{2}{9\pi} } (1 + {\cal M}^2)^{5/2} {\cal M}
\exp \left(-\case{1}{2} {\cal M}^2/\mu^2 \right) d\ell~ .
\end{equation}
\vspace{4.5cm}
\PSbox{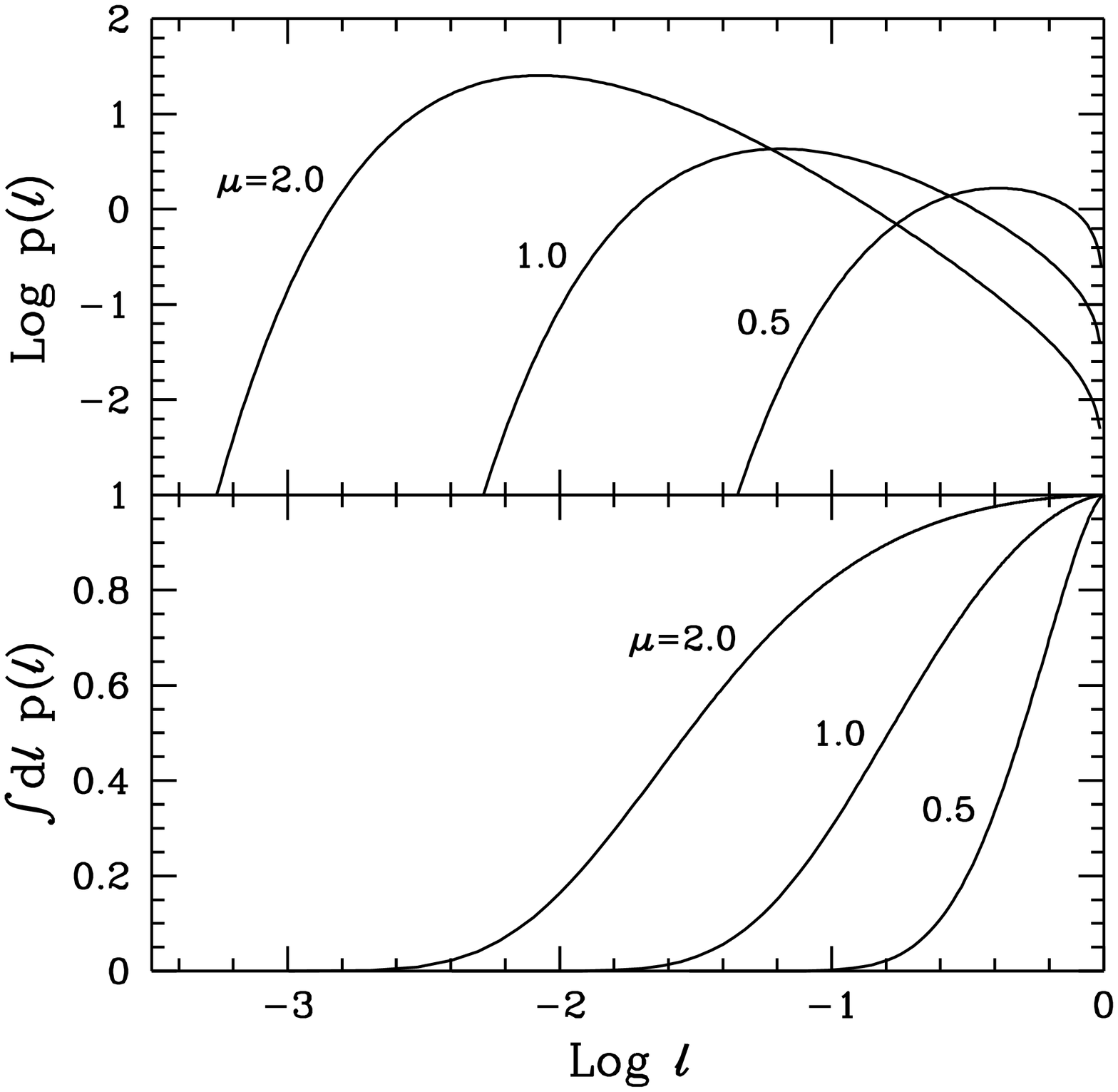 hoffset=-15 voffset=-60 hscale=45 vscale=45}{2.in}{2.in}
\figcaption{The luminosity function (top panel) and the cumulative distribution (bottom panel)
are shown for three different values of the characteristic Mach number, $\mu = \{0.5, 1.0, 2.0\}$.}
\vspace{7mm}

In Figure 1 we plot the LF and the corresponding cumulative distribution
for three plausible values of the characteristic Mach number, $\mu$.  
For the values of $\mu$ shown in Fig. 1, the LF peaks at $\ell \gtrsim 10^{-2}$ and then drops
off rather sharply for decreasing $\ell$.  Furthermore, note that the LF is not well-fit by a power-law
over more than one decade in luminosity.  Therefore, if we suppose that the bright end of the LF is being
observationally selected, then this model shows that a rising observed LF for decreasing luminosity does 
not necessarily mean that we are seeing the ``tip of the iceberg,'' 
as inferred by extrapolating a power-law LF.


\section{DISCUSSION}\label{sec:dis}

In deriving the LF (eq. [\ref{eq:lf}]) using the standard BHL formula for the accretion
rate (eq. [\ref{eq:mdot}]), we have made some rather restrictive assumptions.
The simple BHL formalism 
assumes that the flow is hydrodynamic (i.e., collisional), that the gas density and temperature are 
uniform well beyond the accretion radius of the NS, and that the interaction between the flow and the
magnetic field of the NS can be neglected.  Each of these simplifying assumptions is carried over into the 
derivation of the LF.  A more realistic treatment of the accretion process is beyond the scope of this paper, 
but we can discuss certain phenomenological consequences of lifting some of the aforementioned restrictions.

Roughly speaking, the gas will accrete hydrodynamically if the mean free path of
the atomic or molecular constituents of the gas is shorter than the BHL accretion radius
\citep{begelman77,alcock80}, 
$R_{\rm acc} = 2GM_{\rm ns}V^{-2} \sim 3\times10^{14}~(M / M_\odot) (V / 10 \kms)^{-2} \cm$, where $V$ is the
resultant speed in eq. (\ref{eq:acc-lum}).  For neutral hydrogen, the mean free path is $\sim 10^{16}~n ^{-1} \cm$.  
So, if the ambient gas is predominantly neutral, then a density $\gtrsim 100 \cm^{-3}$ is required for 
hydrodynamic accretion.  On the other hand, the mean free path of a proton in an ionized gas is orders 
of magnitude shorter than the atomic mean free path for $T \sim 10^4~{\rm K}$ \citep{alcock80}.  
Therefore, for globular clusters where the gas is unionized, the BHL accretion process might not be relevant.  
For simplicity, we restrict our discussion to those globular clusters where it is likely that the gas near 
the cluster center is photoionized, 47 Tuc being a case in point \citep[see][]{oconnell97}.

The accretion flow onto a rotating, magnetized NS will differ substantially from the case where
the NS is nonmagnetic.  At the very least, it is expected that the flow will be inhibited
in the former case, either due to a relativistic pulsar wind or something akin to the ``propeller'' 
mechanism \citep{illarionov75}.  It is likely that a detailed consideration of the influence of the 
magnetic field on the infalling plasma, coupled with distributions in field strength and rotation 
frequency, will yield a broader LF than the one shown in Fig. 1.  

There are essentially two factors 
which determine how much gas is present in a globular cluster at the current epoch: (i) the mass and 
central escape speed of the cluster, and (ii) the orbit of the cluster and its current position and 
velocity. Massive globular clusters with a large central escape speed will tend to accumulate more 
gas between crossings through the  Galactic disk.  Also, a cluster will likely  
contain more gas at the current epoch if it is near the top of, or on the descent from, a moderately 
high-altitude Galactic orbit. 

If the gas density is highly inhomogeneous, e.g., strongly enhanced around a discrete number of stars with
large mass-loss rates, the LF derived in \S \ref{sec:lum} would not be valid.  On the other hand, a more clumpy 
gas distribution may help to explain the presence of DCXSs at more than a few core radii from the center of 
their host cluster \citep[see][]{johnston96b}, where the density of an otherwise smooth background of gas 
should be markedly reduced below its central value \citep{vandenberg77}.  In this case, only when a NS 
passes within $\lesssim 0.1 \pc$ of a mass-losing AGB star might the density be sufficiently large to yield 
an accretion luminosity $\gtrsim 10^{30} \ergs$.  

Several of the DCXSs show evidence for significant X-ray variability over timescales of $\sim 1 \yr$
\citep{verbunt98,verbunt00}.  Within the IACN model, such time variability may arise due to spatial
nonuniformity in the density or temperature of the ambient gas over length scales $\lesssim 2~{\rm AU}$ 
(for a NS speed of $10 \kms$).  Hydrodynamic instabilities, possibly induced by the NS magnetic 
field, may lead to variability on much shorter timescales.  Large uncertainties associated with each of these
mechanisms prevent us from making any clear 
predictions for the time variability of the X-ray luminosity of the IACNs.  However, we note that numerical 
simulations of BHL accretion show a strong tendency toward nonsteady behavior \citep[see][]{benensohn97}.   
In contrast, there is no obvious reason why an isolated rotation-powered pulsar should show any 
significant variability in the mean X-ray intensity (averaged over possible X-ray pulsations), 
which may detract from the millisecond pulsar hypothesis if many of the DCXSs are shown to be 
single and variable.


\section{SUMMARY AND CONCLUSIONS}\label{sec:con}

The model we have presented for the DCXSs is rather generic, since the requirements
of the model should not be difficult to satisfy, at least in some globular clusters.
If it turns out that the IACN model explains a significant fraction of the DCXSs,
then much can be learned about the population of single NSs in globular clusters, the properties 
of the intracluster gas, and the BHL accretion process.  On the other hand, if very few of the DCXSs 
are explained by this model (perhaps because most of the DCXSs are shown to be millisecond pulsars or 
binaries), then strong constraints can be placed on the gas content of globular clusters and/or the 
number of single NSs able to accrete efficiently from the intracluster gas.

We list a number of key points regarding the BHL accretion scenario and the IACN population in
globular clusters.

1. Globular clusters with large central escape speeds which are presently high above the Galactic plane 
should contain the largest proportion of IACNs (see \S \ref{sec:dis}).  The globular cluster 47 Tuc is 
a prime candidate in this regard.  The central escape speed and height above the Galactic disk for 47 Tuc 
are $\sim 60 \kms$ and $\sim 3.2 \kpc$, respectively. We also note that there is weak observational 
evidence for the presence of intracluster gas in 47 Tuc, based upon diffuse UV emission from the central 
region of the cluster \citep{oconnell97}, as well as diffuse soft X-ray emission from a possible bow shock 
resulting from the interaction between the cluster gas and the low-density Galactic halo gas 
\citep{krockenberger95}.  In addition, we note that the dispersion measures for 20 of the millisecond
pulsars in 47 Tuc exhibit a range of $\pm 0.2 \cm^{-3} \pc$ around a central value of 
$\simeq 25.5 \cm^{-3} \pc$ \citep{freire00}.  Given that these pulsars are distributed within the inner
$\sim 1 \pc$ of the cluster, and if we assume that the pulsars are spread along the line of sight
by approximately this same amount, this suggests that the electron density within the central region of the 
cluster could be $\sim 0.2 \cm^{-3}$.   

2. If the accreted gas is thermalized at the surface of the NS, we would expect a blackbody 
spectrum, with temperature $kT \sim 50~(L/10^{32} \ergs)^{1/4} f^{-1/4} \ev$, where $f$ 
is the fraction of the NS surface onto which material is accreted.  For a 50 eV black body, the apparent
$B$ magnitude is $\sim 32$ at a distance of 2 kpc.  Therefore, if the accreting NSs radiate as 
blackbodies, and if a negligible amount of the soft X-ray and UV radiation is reprocessed into optical light, 
then these objects should not have detectable optical counterparts.  

3. Theoretical studies indicate that the spectra of the IACNs should be nearly blackbody 
\citep[see][]{alme73,zane00}.  However, based on how poorly understood are the X-ray spectra from better 
studied, more luminous accreting NSs, we would not be surprised if the X-ray spectra of actual BHL 
accreting NSs turned out to differ markedly from Planckian.   

4. We would like to stress that if the IACN model we propose accounts for a very small fraction of the DCXSs, 
then rather stringent constraints may be placed on the properties of the intracluster gas and/or the 
number of NSs able to accrete.  For instance, if we assume a gas temperature of $10^4~{\rm K}$, 
then the maximum accretion luminosity of a nonmagnetic NS is $\lx \sim 10^{32}n \ergs$ 
(eq. [\ref{eq:acc-lum}]).  If no IACNs are seen above the detection threshold, 
$\lx \gtrsim 10^{30} \ergs$, then the gas density must be $n \lesssim 10^{-2} \cm^{-3}$, 
potentially making this the most sensitive available test for the presence of intracluster gas.  

5. If accretion onto magnetized, rapidly rotating, NSs 
is strongly inhibited, then an absence of IACNs could also imply that recycled pulsars dominate the 
population of single NSs.  This is certainly not an untenable hypothesis, since it is likely that a 
NS will undergo at least one mass and angular momentum accretion episode in a binary system over the 
lifetime of the cluster (Rasio, Pfahl, \& Rappaport 2000; Pfahl, Rappaport, \& Rasio 2000, in preparation).

6. A strong density enhancement accompanies the large mass-loss rates and slow wind speeds from AGB stars.
This suggests that some DCXSs may by spatially correlated with AGB stars if the IACN model is successful.
Therefore, it may be worthwhile to calculate a projected 2D correlation function for DCXSs and AGB stars 
for those globular clusters that have been observed with both {\em Chandra} and {\em HST}.

7. Finally, as an aside, we note that the $M^{3}$ dependence of the BHL accretion luminosity 
(see eqs. [\ref{eq:mdot}] and [\ref{eq:acc-lum}]) implies that a black hole of mass $\sim 10 \msun$ 
could be as much as $10^3$ times more luminous that an accreting NS for the same ambient gas
density and temperature \citep[see][]{grindlay78}.  There are theoretical and observational reasons
to believe that there are very few such objects in globular clusters \citep[see][]{sigurdsson93};
however, those that remain might be easily detectable near the cluster centers even if they are not 
in binary systems.


\acknowledgements

This work was supported in part by NASA ATP grant NAG5-8368.  We are grateful to Al Levine
for a careful reading of the manuscript and helpful comments, and to Fred Rasio
for many stimulating discussions about globular clusters.  We also thank the anonymous referee
for bringing point 7 in \S \ref{sec:con} to our attention.



\end{document}